\documentclass{pasj01}
\draft

\begin{document} 
\Received{}
\Accepted{2021/01/08}
\Published{}

\title{A Three-Dimensional Velocity of an Erupting Prominence Prior to a Coronal Mass Ejection}

%%% begin:list of authors
\author{Maria V. \textsc{gutierrez}\altaffilmark{1,2$^{*}$}}
\email{mvgutierreze@iafe.uba.ar}

\author{Kenichi \textsc{otsuji}\altaffilmark{3,4}}
\author{Ayumi \textsc{asai}\altaffilmark{4}}
\author{Raul \textsc{terrazas}\altaffilmark{5}}
\author{Mutsumi \textsc{ishitsuka}\altaffilmark{2}}
\author{Jose \textsc{ishitsuka}\altaffilmark{2$^{\dag}$,6}}
\author{Naoki \textsc{nakamura}\altaffilmark{4}}
\author{Yusuke \textsc{yoshinaga}\altaffilmark{7}}
\author{Satoshi \textsc{morita}\altaffilmark{8}}
\author{Takako T. \textsc{ishii}\altaffilmark{4}}
\author{Satoru \textsc{ueno}\altaffilmark{4}}
\author{Reizaburo \textsc{kitai}\altaffilmark{9,10,4}}
\author{Kazunari \textsc{shibata}\altaffilmark{4,10}}
\footnotetext[$^\dag$]{This is the former affiliation.}

\altaffiltext{1}{Institute for Research in Astronomy and Astrophysics (IAFE CONICET-UBA), 
Buenos Aires, Argentina}
\altaffiltext{2}{Geophysical Institute of Peru, Lima, Peru}
\altaffiltext{3}{National Institute of Information and Communications Technology (NICT), 
Koganei, Tokyo, 184-8795, Japan}
\altaffiltext{4}{Astronomical Observatory, Kyoto University,
Sakyo, Kyoto, 606-8502, Japan}
\altaffiltext{5}{Ica Solar Station, Department of Physics, 
San Luis Gonzaga National University of Ica, Ica, Peru}
\altaffiltext{6}{National University of the Center of Peru, Huancayo, Peru}
\altaffiltext{7}{Department of Astronomy, Kyoto University,
Sakyo, Kyoto, 606-8502, Japan}
\altaffiltext{8}{National Astronomical Observatory of Japan,
Osawa, Mitaka, Tokyo, 181-8588, Japan}
\altaffiltext{9}{Bukkyo University, Kita, Kyoto, 603-8301, Japan}
\altaffiltext{10}{Ritsumeikan University, Kita, Kyoto, 603-8577, Japan}
\altaffiltext{11}{Doshisha University, Kyotanabe, Kyoto, 610-0394, Japan}
%%% end:list of authors


\maketitle

%%%%%%%%%%%%%%%%%%%%%%%%%%%%%%%%%%%%%%%%%%%%%%%%%%%%%%%%%%%%%%%%%%%%%%
\begin{abstract}

We present a detailed three-dimensional (3D) view of a prominence eruption, coronal loop expansion,
and coronal mass ejections (CMEs) associated with an M4.4 flare that occurred on 2011 March 8 
in the active region NOAA 11165.
Full-disk H$\alpha$ images of the flare and filament ejection were successfully obtained 
by the Flare Monitoring Telescope (FMT) following its relocation to Ica University, Peru. 
Multiwavelength observation around the H$\alpha$ line enabled us to derive 
the 3D velocity field of the H$\alpha$ prominence eruption. 
Features in extreme ultraviolet were also obtained by the Atmospheric Imager 
Assembly onboard the {\it Solar Dynamic Observatory} and 
the Extreme Ultraviolet Imager on board the {\it Solar Terrestrial Relations 
Observatory - Ahead} satellite. 
We found that, following collision of the erupted filament with the coronal magnetic field, 
some coronal loops began to expand, leading to the growth of a clear CME. 
We also discuss the succeeding activities of CME driven by multiple interactions 
between the expanding loops and the surrounding coronal magnetic field.

\end{abstract}

%%%%%%%%%%%%%%%%%%%%%%%%%%%%%%%%%%%%%%%%%%%%%%%%%%%%%%%%%%%%%%%%%%%%%%
\section{Introduction} \label{sec:intro}

Solar flares are explosive events occurring in the solar corona 
that release extremely large quantities ($10^{28}$--$10^{32}$~erg) 
in radiative, kinetic, thermal and non-thermal form.
They are considered to be one of the most attractive scientific objects in solar physics, with
emissions across the electromagnetic spectrum originating from atmospheric layers extending 
from the chromosphere, in extreme cases the photosphere, to the outer regions of the solar corona. 
Solar flares can also produce high-speed coronal mass ejections (CMEs) 
that travel through interplanetary space with occasional serious impacts on the solar-terrestrial 
environment. 

A filament, relatively cool, dens plasma supported by magnetic field and floated 
in the corona, sometimes erupts and cause or be part of a CME.
Therefore, the relations among filament eruptions, flares and CMEs have been earnestly studied (e.g., \cite{Schmieder2013,Schmieder2015,Schmieder2020}).
\citet{Munro1979} reported that more than 70~\% of CMEs are associated with eruptive
prominences of filament disappearance, both with and without large soft X-ray (SXR) 
enhancement, that are accordingly recognized as flares.
\citet{Gopal2003} also reported that 72~\% of prominence eruptions observed in microwaves 
are associated with CMEs.
Filaments eruptions are thought to be caused by loss of equilibrium and/or magnetohydrodynamic instability \citep{Demoulin2010,Kliem2006,Mackay2010}.

Finding the origin of a CME in the solar lower atmosphere and following its evolution to be a CME
are, however, sometimes difficult owing to a gap between the coronagraphic observational field-of-view 
of the CME and that for flares and filament eruptions on the solar surface. 
The flare-CME connection is a controversial topic and of great interest 
to the scientific community, which has not yet found an established relationship 
between these phenomena.
%%%
Some flares, referred to as confined flares, are not eruptive and not associated with CMEs 
\citep{Moore2001,Mackay2010}.
Some CMEs are not related with flares, which are defined as strong X-ray emission, but related 
with eruption of quiet region filaments and/or giant arcade formation seen in X-rays \citep{McAllister}.
Therefore, the connection between a CME and the related phenomena on the solar surface should be  
cleared more by focusing on the temporal evolution of the flare and its relation to the CME. 
Accordingly, in this study we investigate the initial phases of CME-related filament eruptions 
by using H$\alpha$ data with the goal of detecting signatures preceding the launch of CMEs.
Recently, a three-dimensional (3D) prominence reconstruction was done by using IRIS data \citep{Schmieder2017}.
The authors succeeded to demonstrate that an helical structure of the prominence seen 
in the IRIS images consists in fact of horizontal field lines parallel to the solar surface.
Spectroscopic observations of eruptive filaments have become more important 
to reveal the relation between CMEs and filament eruptions.

To this end, we examined a set of solar phenomena that occurred on 2011 March 8, 
in Active Region (AR) NOAA 11165.
We examined the erupted prominence in detail and followed its 3D temporal evolution and relation 
to the overlying coronal magnetic field.
The remainder of this paper is organized as follows.
We describe the observations in Section~2 and 
present analysis and results on the prominence eruption and coronal loops prior to the appearance 
of the CMEs in Sections~3 and 4, respectively.
Finally, a summary and discussion are given in Section~5.

%%%%%%%%%%%%%%%%%%%%%%%%%%%%%%%%%%%%%%%%%%%%%%%%%%%%%%%%%%%%%%%%%%%%%%
\section{Observations}

\begin{figure}
 \begin{center}
  \includegraphics[width=12cm]{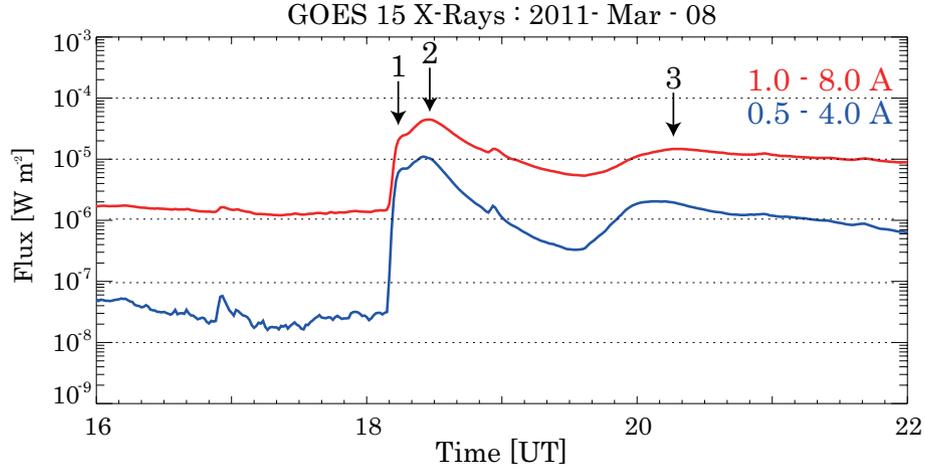} 
 \end{center}
\caption{{\it GOES} X-ray emission at 1.0 -- 8.0 {\AA} (red) and 0.5 -- 4.0 {\AA} (blue) channels 
of the 2011 March 8 flare.
Arrow 2 indicates the peak of the M4.4 flare at 18:28~UT.
The peak at 20:16~UT corresponds to an M1.5 flare that occurred in the same AR NOAA~11165 
(shown with arrow 3).}
\label{fig1}
\end{figure}

On 2011 March 8, an M4.4 solar flare with a soft X-ray (SXR) classification was recorded at AR 
NOAA 11165, which is located near the southwest solar limb (S17$^{\circ}$, W88$^{\circ}$),
by the {\it Geostationary Operational Environmental Satellite} ({\it GOES}). 
Figure~\ref{fig1} shows the SXR light curves taken by {\it GOES}.
The flare began at 18:08~UT and peaked at 18:28~UT (shown in the figure with arrow 2), 
with light curves featuring several small humps and subpeaks.
This flare was followed by another gradual M1.5 flare that began and peaked at 19:35 and 20:16~UT 
(arrow 3), respectively.

%%%%%%%%%%%%%%%%%%%%%%%%%%%%%%%%%%%%%%%
%%%%%%%%%%%%%%%%%%%%%%%%%%%%%%%%%%%%%%%

A filament/prominence eruption associated with the flare can be seen in H$\alpha$ 
images obtained by the Flare Monitoring Telescope (FMT; \cite{Kurokawa}),
which was originally installed at Hida Observatory, Kyoto University, Japan, and 
relocated to Ica University, Peru in 2010 under the international collaboration 
of the Continuous H-Alpha Imaging Network (CHAIN) Project \citep{UeNo2007}.
FMT provides full-disk images in five wavelengths: 
H$\alpha$ line center (6562.8~{\AA}); wings at $+0.8$~{\AA} and $-0.8$~{\AA}; 
continuum (6100~{\AA}) images; and another H$\alpha$ line center image 
to detect limb prominences. 
The spatial and temporal samplings of FMT are about $2^{\prime\prime}.0$, 
and 20~s, respectively.
In this paper, we will use the term `FMT-Peru' to clarify that the data we used in this study
were obtained following the relocation.

\begin{figure}
 \begin{center}
  \includegraphics[width=14cm]{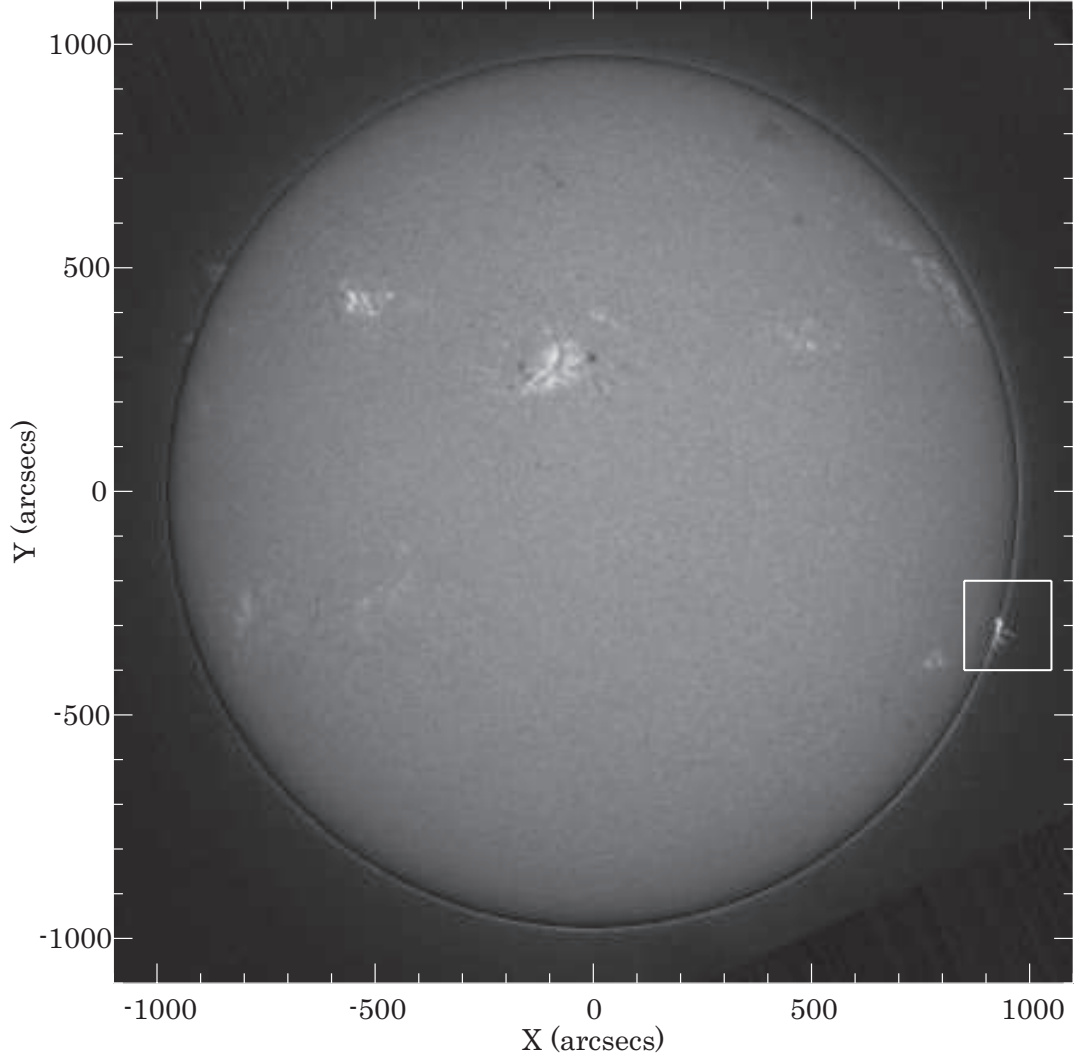} 
 \end{center}
\caption{Full-disk solar image (solar north is up and west is to the right) of the flare 
in H$\alpha$ center taken at 18:14~UT by FMT-Peru. 
The white box shows the flaring region of AR NOAA~11165.}
\label{fig2}
\end{figure}

The telescope is capable of measuring the 3D velocity fields of 
active filaments/prominences \citep{Morimoto2003a,Morimoto2003b,Denis2017}, 
and Moreton waves \citep{Eto2002,Narukage2002,Narukage2008,Asai2012,Denis2019}.
Figure~\ref{fig2} shows a full-disk H$\alpha$ center image of the flare taken by FMT-Peru, 
with the erupting feature visible in the region marked with the white box on the southwest limb.

%%%%%%%%%%%%%%%%%%%%%%%%%%%%%%%%%%%%%%%
%%%%%%%%%%%%%%%%%%%%%%%%%%%%%%%%%%%%%%%

\begin{figure}
 \begin{center}
  \includegraphics[width=12cm]{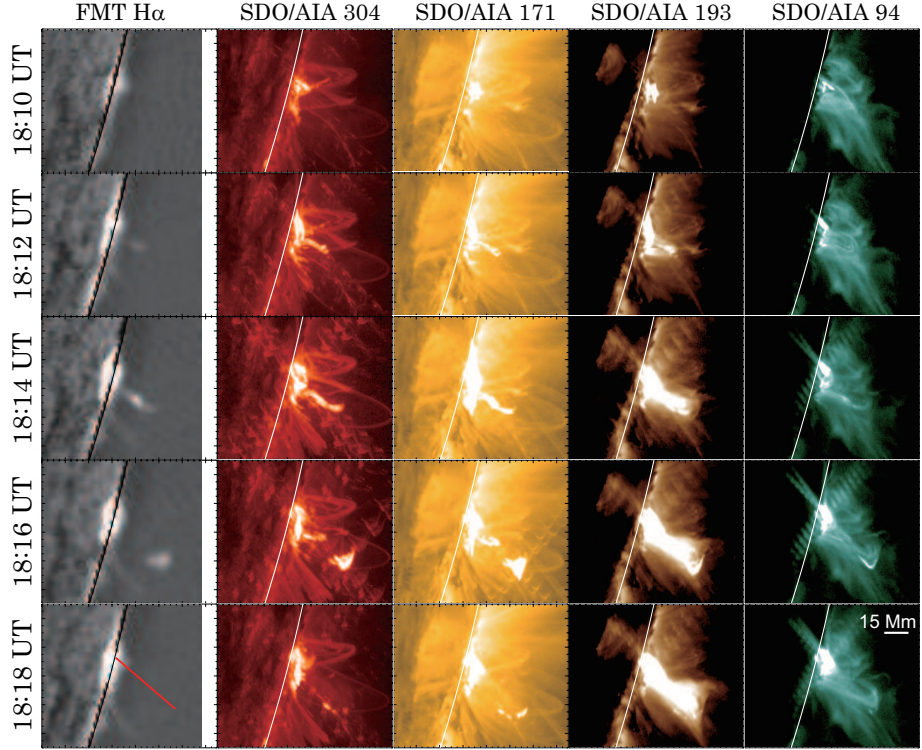} 
 \end{center}
\caption{Multiwavelength time sequences of the 2011 March 8 flare.
Columns from left to right are FMT-Peru H$\alpha$ combined, and 
{\it SDO}/AIA at 304, 171, 193, and 94~{\AA} images, respectively.
The red solid line in the bottom left panel indicates the slit position used in Figure~\ref{fig7}.}
%See also the supplementary movie 1 for the temporal evolution of the H$\alpha$ images.}
\label{fig3}
\end{figure}

To view the spatial distribution of prominence gases irrespective to line-of-sight motion,
we created `combined' images of three wavelengths (center and $\pm0.8$~{\AA}).
Prominences/filaments moving rapidly in the line-of-sight direction are often invisible
with images of only one wavelength due to the Doppler shift of the H$\alpha$ line.
The images of three wavelengths were, therefore, combined into time series to enable tracking 
of the evolution of overall prominence gas.
%%%
At each wavelength, a specific intensity level was set and the brighter regions above that level 
were extracted to reveal the prominences involved in the selected region.
Finally, we superimposed the images extracted at the three wavelengths 
at each time step to produce a time sequence of combined H$\alpha$ images,
as shown in the leftmost column of Figure~\ref{fig3}.
%%%
See also the supplementary movie 1 for the temporal evolution of the H$\alpha$ images.
It shows images at H$\alpha$ $-0.8$~{\AA} (top left), center (top right), $+0.8$~{\AA} (bottom left), 
respectively.
The bottom right panel is the combined images.

%%%%%%%%%%%%%%%%%%%%%%%%%%%%%%%%%%%%%%%
%%%%%%%%%%%%%%%%%%%%%%%%%%%%%%%%%%%%%%%

Flare and eruption features were also captured in extreme ultraviolet (EUV) 
images taken by the Atmospheric Imaging Assembly (AIA; \cite{Lemen2012}) 
onboard the {\it Solar Dynamics Observatory} ({\it SDO}; \cite{Pesnell2012}).
AIA takes full-disk solar images with a temporal resolution of 12~s, 
while the images of filters 1600 and 1700~{\AA} are taken every 24~s. 
The pixel size of the images is $0^{\prime\prime}.6$.
In this study, we used images taken with 304, 171, 193, and 94~{\AA} filter,
which are mainly attributable to the He~{\sc ii} (80,000K), Fe~{\sc ix} (0.6MK), 
Fe~{\sc xii} (1.2MK),  and Fe~{\sc xviii} lines (6.3MK), respectively.
The second to fifth columns from the left in Figure~\ref{fig3} show the temporal evolution 
of the flare and the prominence eruption at 304, 171, 193, and 94~{\AA}, respectively.

The prominence eruption associated with the flare is seen as a bright feature
in the H$\alpha$ images that moves to the southwest from the flare site.
It is also seen as bright features (moving blobs) in the AIA 304 and 171~{\AA} images.
In particular, common features are seen on the erupting blobs in the H$\alpha$, 
EUV 304 and 171~{\AA} images.
By contrast,in the AIA 193 and 94~{\AA} images outer coronal loops with higher temperature 
are predominant.
As is often observed \citep{Chifor2006}, the erupting filament suffers heating 
at the outer edge and is therefore visible even in EUV bands such as 171 and 193~{\AA}, 
which are sensitive to higher temperatures.
%%%
Here, we have to note that the AIA channels have broad temperature response.
It is often difficult to know whether a structure is really hot or cool.
After 18:16~UT, the prominence becomes unclear in both H$\alpha$ and 304~{\AA},
while the coronal loops visible in 193 and 94~{\AA} are still evolving and brightening.

The Extreme-Ultraviolet Imager (EUVI; \cite{Wuelser}), part of the Sun-Earth
Connection Corona and Heliospheric Investigation (SECCHI; \cite{Howard2008}) 
onboard the {\it Solar Terrestrial Relations Observatory - Ahead} 
({\it STEREO-A}; \cite{Kaiser2008}), also observed the flare and prominence 
eruption (filament eruption in the view of {\it STEREO-A}) in the EUV bands.
At the time of observation, {\it STEREO-A} was located $87.^{\circ}6$ ahead of the Earth
on a heliocentric elliptical orbit in the ecliptic plane and, therefore,
was able to capture a top view of the flaring region (NOAA 11165).
We used 171~{\AA} EUVI images in which the emission line of Fe~{\sc ix} (0.6MK) was 
dominant to compare the temporal evolution of the erupting blobs.

The flare and prominence eruption were followed by a CME, which was first seen 
at 19:00 UT in the field of view of coronagraph C2 on the Large Angle and Spectrometric 
COronagraph (LASCO; \cite{Brueckner}) onboard {\it SOHO}.
We will discuss this CME further in Section~4.

To examine the temporal evolution and dynamics of the flare and prominence eruption,
we used hard X-ray (HXR) data obtained by the {\it Reuven Ramaty 
High Energy Solar Spectroscopic Imager} ({\it RHESSI}; \cite{Lin}) and
microwave data taken at the Sagamore Hill Solar Observatory under 
the Radio Solar Telescope Network (RSTN; \cite{Guidice}).
We focused in particular on 25 -- 50 and 50 -- 100~keV HXR emissions measured by 
{\it RHESSI} and microwave emissions at the 4.99, 8.8, and 15.4~GHz channels 
measured by RSTN.
At these wavelengths (in this case given, respectively, as energy bands and frequencies)
it is possible to observe nonthermal Bremsstrahlung and gyrosynchrotron emissions 
associated with accelerated electrons, allowing for the examination of the energetic features 
of accelerated electrons in the flare.

%%%%%%%%%%%%%%%%%%%%%%%%%%%%%%%%%%%%%%%%%%%%%%%%%%%%%%%%%%%%%%%%%%%%%%
\section{3D Velocity Field of the Prominence Eruption}

There have been some reports on the derivation of the Doppler (line-of-sight) velocities
of H$\alpha$ disk filaments observed using FMT \citep{Morimoto2003a,Morimoto2003b,Denis2017}.
The methods used in these studies are based on the `cloud model', which was originally 
suggested by \citet{Beckers} and \citet{Mein} and modified for FMT data.
By contrast, in this study we observed an off-limb bright prominence with a dark background, 
which meant that we could not use the cloud-model-based methods.

Instead, to derive Doppler velocities we assumed that the observed profiles were emitted 
from a slab with a uniform and constant source function.
Based on the three wavelength data points (H$\alpha$ center and $\pm0.8$~{\AA}) 
and an additional assumption that the H$\alpha$ intensities at the $\pm4.0$~{\AA}
wings are equal to zero, profile fitting produced four unknown parameters 
(i.e., the source function, optical depth, Doppler velocity, and Doppler width) with an error 
of about $\pm10$~km~s$^{-1}$.
Owing to misalignment of the wavelengths of Peru-FMT filters from their nominal values, 
the actual error in determined Doppler velocity was potentially worse at roughly $\pm15$~km~s$^{-1}$.
Furthermore, FMT-Peru cannot detect blobs with velocities greater than 50~km~s$^{-1}$.
%%%
We, in this paper, fitted the erupting filament only with images of the H$\alpha$ center 
and the wings with the band position of $\pm$~0.8~{\AA} and with the passband width of about 0.5~{\AA}.
Profiles of prominences observed with spectrographs often show a narrow H$\alpha$ line 
with a maximum FWHM between 0.6 and 0.8~{\AA} \citep{Ruan}.
In our case the wide bandpass of the filters around 0.5~{\AA} allows us to integrate 
the prominence emission from the inflexion point of the line profile (0.45~{\AA}) to the far wing.
%%%%%
Therefore, the prominences are visible in our three filters, while there is a large uncertainty 
in the Doppler shift values.
%%%%%

\begin{figure}
 \begin{center}
  \includegraphics[width=10cm]{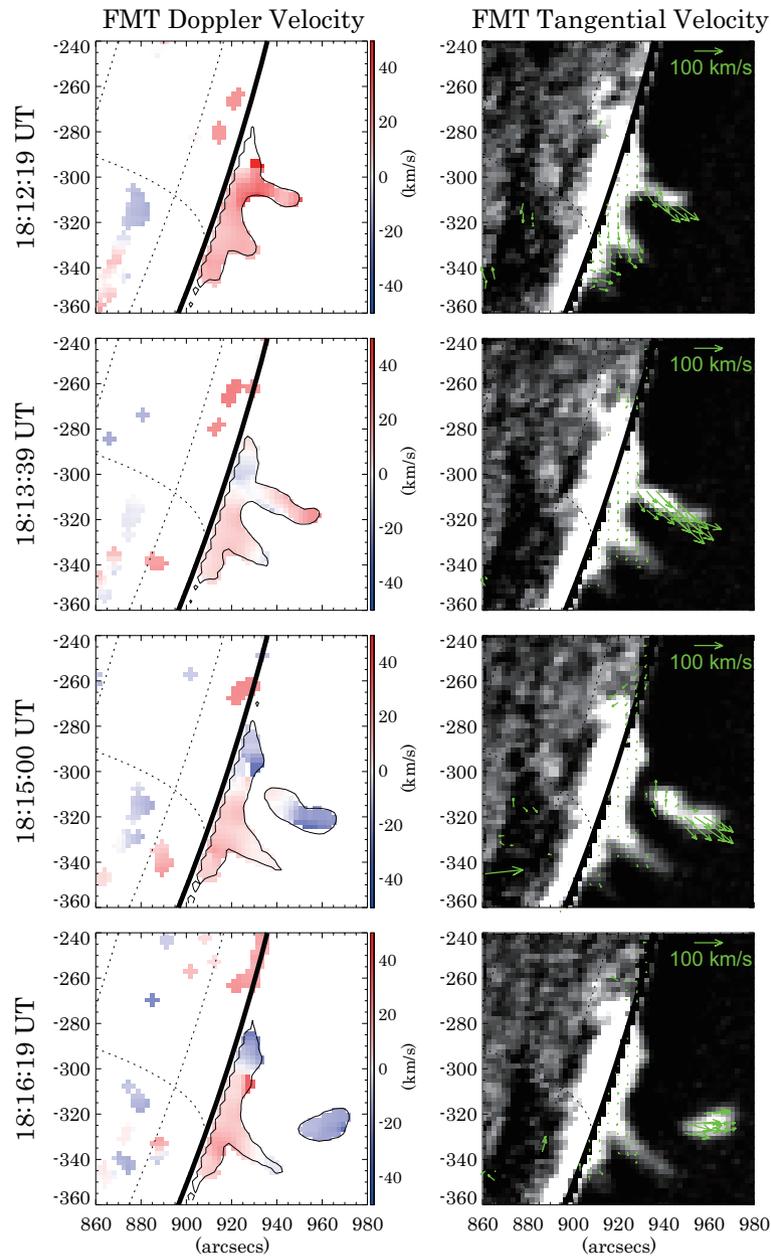} 
 \end{center}
\caption{Temporal evolution of 3D velocity of erupting prominence. 
{\it Left}: Doppler velocity shown with blue- and red-shifted features colored 
blue and red, respectively.
{\it Right}: Tangential velocity with direction and magnitude indicated with green arrows.
The background images are FMT-Peru H$\alpha$ combined images.
The sign of the Doppler shift changes from red to blue at around 18:14~UT; 
at the same time, the direction of the tangential velocity changes 
from southwest to slightly northwest.}
\label{fig4}
\end{figure}

The left panels of Figure~\ref{fig4} show temporal evolution maps of Doppler velocity 
with blue and red colors indicating blue- and red-shifted Doppler velocities, respectively.
The erupting blob initially has a red-shift with a velocity of about 25~km~s$^{-1}$ 
that changes at around 18:14~UT to a blue-shift with a velocity of about 30~km~s$^{-1}$.
This timing (18:14~UT) corresponds to arrow 1 in Figure~\ref{fig1}.

By assuming that the observed H$\alpha$ prominence is optically thin, 
it is possible to combine the optical thickness and the source function into one unknown parameter
-- the emissivity -- and reduce the number of unknown physical parameters to three.
We derived the Doppler velocities by fitting the data  under this assumption and then 
confirmed that the resulting velocities were consistent with those derived with the former method.
The Doppler velocities in the remainder of the study were therefore derive using the 
former method.

%%%%%%%%%%%%%%%%%%%%%%%%%%%%%%%%%%%%%%%
%%%%%%%%%%%%%%%%%%%%%%%%%%%%%%%%%%%%%%%

We derived the tangential velocity of the ejected filament 
(i.e., the velocity of the filament in the plane of the sky) by applying the local 
correlation tracking (LCT) method to the H$\alpha$ combined image sequence.
The right panels of Figure~\ref{fig4} show the temporal evolution maps of the 
tangential velocity, with the green arrows showing the direction and amplitude of the tangential velocity.
The tangential velocity of the erupting filament is roughly 80~km~s$^{-1}$ and rapidly decelerating.
The direction of the tangential velocity changes from southwest to slightly northwest
at around 18:14~UT, corresponding temporally to the sudden direction change of the Doppler velocity.

\begin{figure}
 \begin{center}
  \includegraphics[width=8cm]{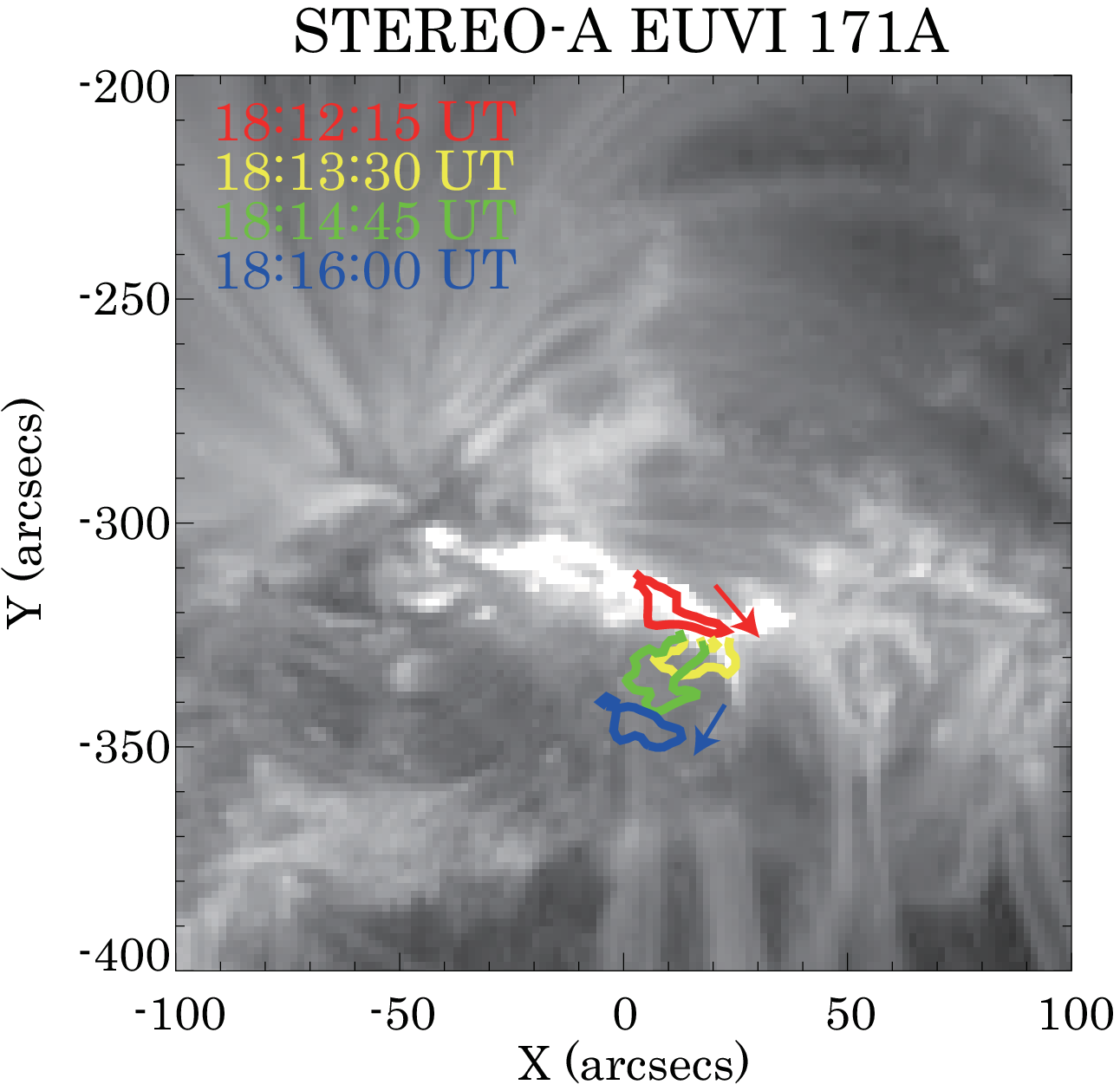} 
 \end{center}
\caption{Temporal evolution of filament eruption seen in the 
{\it STEREO-A}/SECCHI/EUVI 171~{\AA} images.
The contours show the bright feature associated with the erupting filament, with 
different colors indicating different times.
The background is an EUVI 171~{\AA} image taken at 18:11~UT}.
\label{fig5}
\end{figure}

Because the flare site was nearly on the solar limb (W88$^{\circ}$), 
{\it STEREO-A}, which was located $87.^{\circ}6$ ahead of the Earth at the time of
observation, was able to capture a top view of the flare and the filament eruption. 
As discussed above, the erupting H$\alpha$ prominence was co-spatial 
with a bright blob seen in the AIA 171~{\AA} images (see Figure~\ref{fig3}), and
therefore we investigated the temporal evolution of the filament eruption 
in the {\it STEREO-A}/SECCHI/EUVI 171~{\AA} images.
%%%
Figure~\ref{fig5} shows the temporal evolution of the erupting filament 
in the EUVI 171~{\AA} images.
The contours outline a bright feature (bright blob) associated with 
the erupting filament, with different colors indicating different times.
It is seen that the filament predominantly moves southward, with a slight 
westward motion before 18:14~UT, followed by eastward motion afterwards.
These motions correspond to motion away from and toward the Earth, respectively.
The centroid of the contoured region moves at roughly 40~km~s$^{-1}$ between 18:12:15~UT (red) and
18:13:30~UT (yellow) and at 40~km~s$^{-1}$ between 18:14:45~UT (green) and 18:16:00~UT (blue).

%%%%%%%%%%%%%%%%%%%%%%%%%%%%%%%%%%%%%%%
%%%%%%%%%%%%%%%%%%%%%%%%%%%%%%%%%%%%%%%

Analysis of the blob dynamics in the prominence eruption reveal an interesting behavior
in which the blobs are deflected at specific heights within the corona, indicating an interaction 
between the blobs and the coronal environment.
In the next section, we examine the reactions consequential to this interaction.

%%%%%%%%%%%%%%%%%%%%%%%%%%%%%%%%%%%%%%%%%%%%%%%%%%%%%%%%%%%%%%%%%%%%%%
\section{EUV Flux Rope Expansion and CME}

\begin{figure}
 \begin{center}
  \includegraphics[width=14cm]{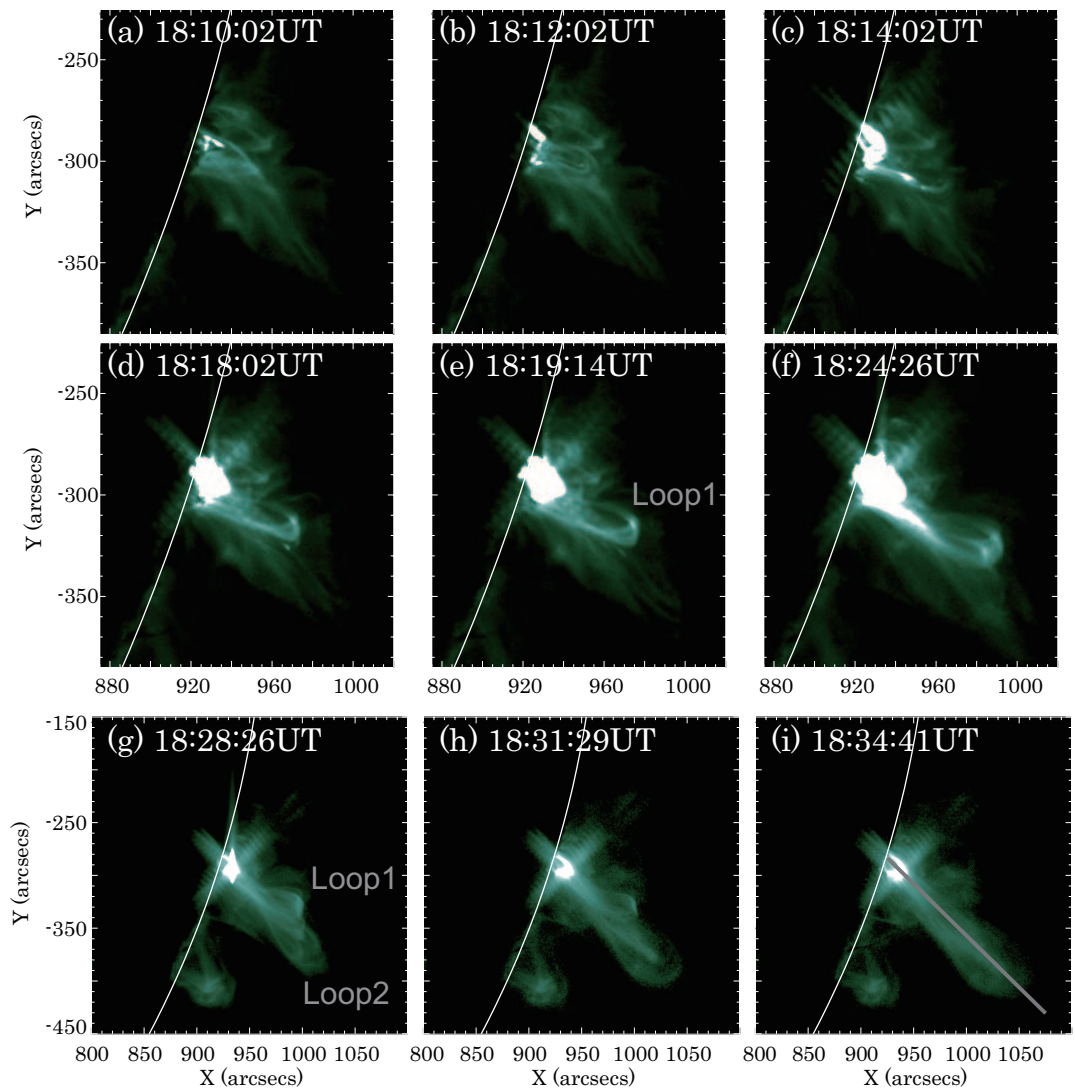} 
 \end{center}
\caption{Temporal evolution of expanding loops seen in the
{\it SDO}/AIA 94~{\AA} images.
Note that the field of view of panels (g) to (i) is wider that from (a) to (f).
The gray solid line in panel (i) indicates the slit position used in Figure~\ref{fig7}.}
%See also the supplementary movie 2 for the temporal evolution.}}
\label{fig6}
\end{figure}

Following the disappearance of the cool H$\alpha$ prominence at around 18:18~UT, 
further activity commences in the higher corona, in the form of expansion of the 
outer coronal loops seen in the SDO/AIA 94~{\AA} band.
This activity appears to lead to a CME.
%%%
Figure~\ref{fig6} shows the temporal evolution of the expanding loops.
From 18:10~UT onward, the expansion of a coronal loop is seen (Loop1).
After 18:19~UT, another loop (Loop2) begins to expand from just south of Loop1.
Loop1 remains in place or expand very slowly, and disappears at 18:35~UT.
On the other hand, Loop2 rapidly expands until it eventually leaves the field of view.

%%%%%%%%%%%%%%%%%%%%%%%%%%%%%%%%%%%%%%%
%%%%%%%%%%%%%%%%%%%%%%%%%%%%%%%%%%%%%%%

\begin{figure}
 \begin{center}
  \includegraphics[width=11cm]{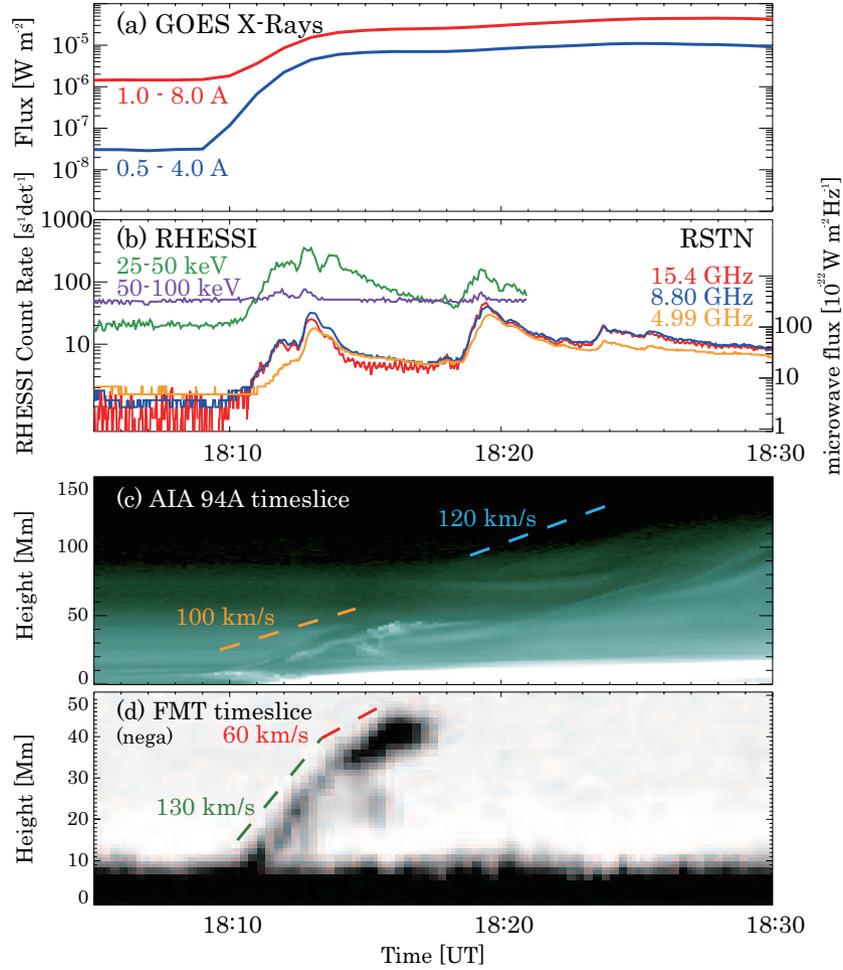} 
 \end{center}
\caption{
(a) Soft X-ray light curves in the {\it GOES} 1.0 -- 8.0 (red) and 0.5 -- 4.0~{\AA} 
(blue) channels.
(b) Hard X-ray light curves measured with {\it RHESSI} at
25 -- 50 (green) and 50 -- 100~keV (purple) and microwave flux taken with 
RSTN in the 4.99 (orange), 8.8 (blue), and 15.4 (red) channels.
(c) Time-sequenced EUV (94~{\AA}) image (time slice diagram) obtained with
{\it SDO}/AIA along the slit line shown in Figure~6(i).
(d) Time slice diagram of H$\alpha$ combined image obtained with FMT-Peru
along the slit line shown in the bottom left panel of Figure~3.
The zeros of the vertical axis for both time-slice diagrams correspond to the lower 
edges of the slit lines.}
\label{fig7}
\end{figure}

To study the temporal evolution of the vertical motion of these activities 
in more detail, we constructed the time slice diagrams (time-sequenced images along 
slit lines) shown in the bottom panels of Figure~\ref{fig7}.
Panels (c) and (d) show time slice diagrams for the coronal loops of
AIA 94~{\AA} (negative images) and the H$\alpha$ prominence, respectively.
The zero of the vertical axis on each panel corresponds to the lower edge of the slit line
(i.e., the edge closer to the solar limb).
The positions of the slits are also shown in Figure~\ref{fig6}(i) and in the bottom left panel 
of Figure~\ref{fig3}.
%%%
As described in Section~3, the H$\alpha$ prominence erupts and decelerates 
after 18:14~UT prior to fading out at around 18:18~UT.
The vertical motion of the EUV 94~{\AA} loop (Loop1) is seen to be associated with the
H$\alpha$ prominence eruption.
Loop1 also shows rapid acceleration and sudden deceleration corresponding to motions 
in the H$\alpha$ prominence eruption.
The Loop1 and H$\alpha$\ prominence ascending velocities of about 130 and 100~km~s$^{-1}$, 
respectively.
After 18:19~UT, another vertical motion associated with Loop2 is seen 
in the time slice diagram of the EUV 94~{\AA} images.
The ascending velocity in this case is about 120~km~$^{-1}$.

Figures~\ref{fig7}(a) and (b) show the light curves at SXRs (1.0 -- 8.0 and 
0.5 -- 4.0~{\AA}) taken by {\it GOES},
HXRs (25 -- 50 and 50 -- 100~keV) taken by {\it RHESSI}, and
microwaves (4.99, 8.8, and 15.4~GHz) taken by RSTN.
%%%
In the HXR and microwave light curves two bursts, with peaks 
at 18:12 -- 18:14 and at 18:19 -- 18:20~UT, are seen.
The short-lived nature of these features and the simultaneity of the emissions 
suggest that both the microwave and HXR emissions are of nonthermal origin.
The timings of these nonthermal bursts are associated with the rapid eruptions
of H$\alpha$ prominence and EUV coronal loops.
The temporal association between the vertical motion (i.e., outward rapid acceleration) 
and the nonthermal emissions indicates that plasmoids are being ejected with the release of 
high amounts of magnetic energy as has previously been reported 
(e.g., \cite{Kahler1988,Ohyama1998,Asai2004,Asai2006,Nishizuka2010,Takasao2016}).
These results support the so-called `plasmoid-induced-reconnection' model
\citep{Shibata1999,Shibata2001}.

%%%%%%%%%%%%%%%%%%%%%%%%%%%%%%%%%%%%%%%
%%%%%%%%%%%%%%%%%%%%%%%%%%%%%%%%%%%%%%%

\begin{figure}
 \begin{center}
  \includegraphics[width=12cm]{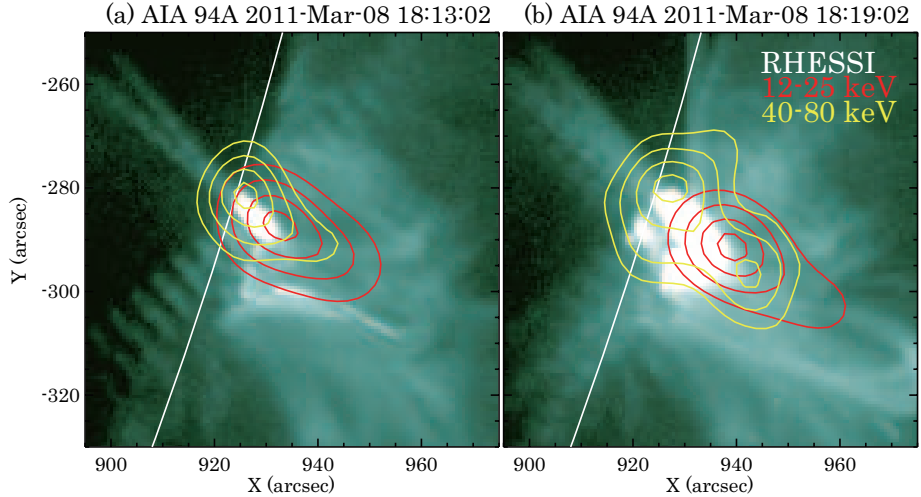} 
 \end{center}
\caption{
Spatial distribution of the emission sources. 
The backgrounds are AIA 94~{\AA} images taken at 18:13:02 (a) and 18:19:02~UT (b).
The contours show the {\it RHESSI} 12 -- 25~keV intensity (red), and the 40 -– 60~keV
intensity (yellow), respectively. 
The contour levels are 40, 60, 80, and 95\% of the peak intensity.}
\label{fig8}
\end{figure}

Figure~\ref{fig8} shows spatial distributions of emission sources in HXR for 
the nonthermal bursts around at 18:13~UR (Fig.~\ref{fig8}a)  and 18:19~UT (Fig.~\ref{fig8}b).
The backgrounds are AIA 94~{\AA} images are taken at 18:13:02 and 18:19:02~UT, respectively.
We overlaid HXR contour images observed with {\it RHESSI} in 12 -- 25 ~keV (red) 
and 40 -- 80~keV (yellow).
The levels of the contours are 40\%, 60\%, 80\%, and 95\% of the peak intensity.
We synthesized the HXR images by using grids 3 -– 8 and integrating
over 40 seconds for the first peak and 80 seconds for the second peak.
We recognize a HXR emission source in the low energy band (12 -- 25~keV)
appear above the flare loops seen in the 94~{\AA} images in the both times.
These emission sources extend in the vertical direction, i.e. along the current sheet.
The HXR emission in the high energy band (40 -- 80~keV), on the other hand, mainly comes from 
the footpoint of the flare loops.
These are consistent with the previous study \citep{Su}.
We can also see a ``loop-top'' HXR emission source \citep{Masuda} or an extension of the contour line 
in the high energy HXR band.
These loop-top HXR emission sources are located just above the top of the bright flare loops 
seen in the 94~{\AA}, and are probably associated with the interaction between the 
reconnection outflow jet and/or plasmoid ejected from the reconnection region and the flare loops
\citep{Takasao2016}. 

%%%%%%%%%%%%%%%%%%%%%%%%%%%%%%%%%%%%%%%
%%%%%%%%%%%%%%%%%%%%%%%%%%%%%%%%%%%%%%%

\begin{figure}
 \begin{center}
  \includegraphics[width=14cm]{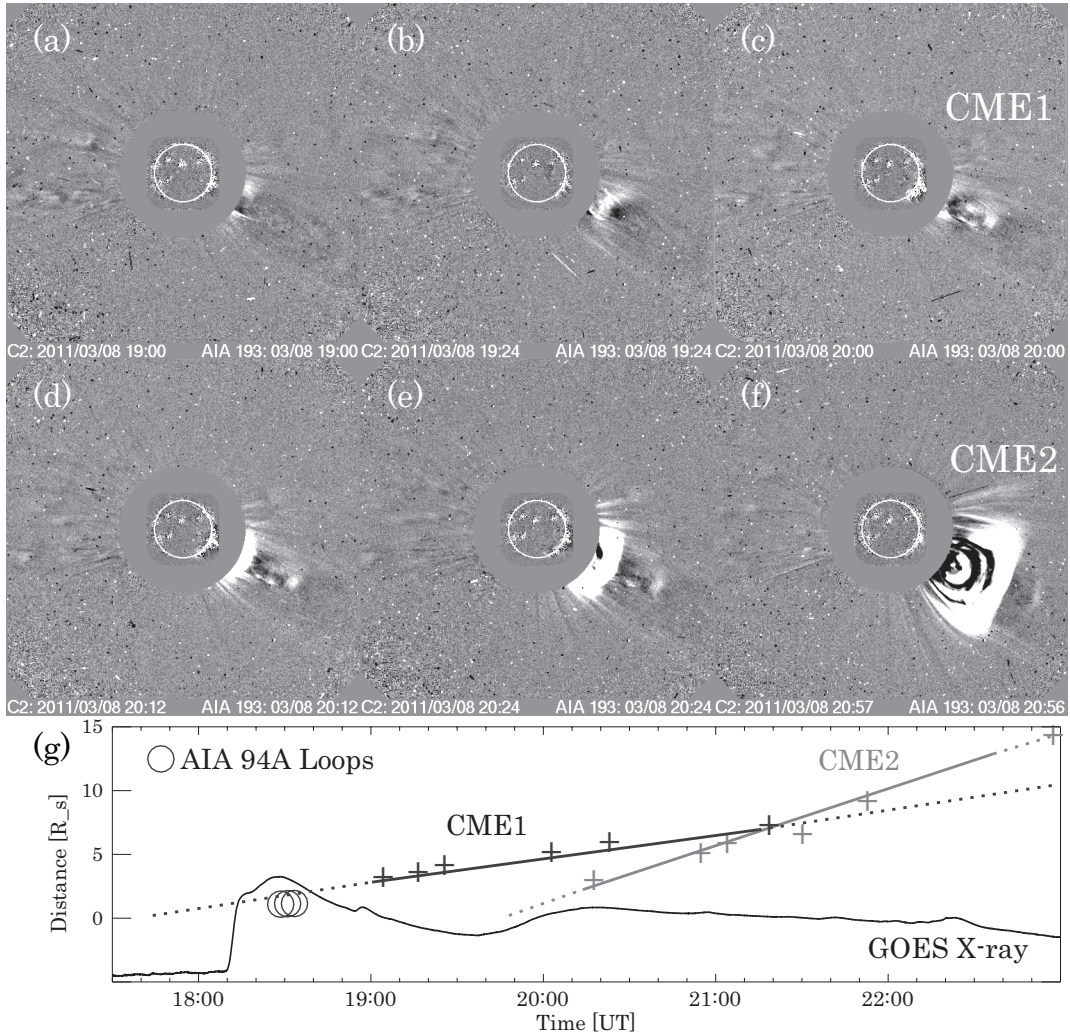} 
 \end{center}
\caption{
(a -- f) Time sequences of {\it SOHO}/LASCO C2 running difference images.
Each image is overlaid with a corresponding EUV 193~{\AA} running difference image obtained 
from {\it SDO}/AIA.
(g) Time-distance diagram of CMEs and EUV coronal loops.
The distance is measured from the solar limb.
The CME front distances for CME1 (CME2) at a given time are represented by black (gray) 
plus $+$ signs, while the circles $\bigcirc$ mark the position of the EUV coronal loop (Loop2).
R$_s$ is the solar radius, ($\approx$ 695,800~km.
For comparison, we also plot the SXR flux in the {\it GOES} 1.0 -- 8.0~{\AA} channel.}
\label{fig9}
\end{figure}

The {\it SOHO}/LASCO C2 coronagraph detected the appearance of a faint and slow CME  
at 19:06~UT with a linear velocity of 280~km~s$^{-1}$
(see the {\it SOHO}/LASCO CME online catalog, http://cdaw.gsfc.nasa.gov/CME\_list/)
\citep{Yashiro2004}. 
From the timing and direction of expansion, this CME appears to be associated with
the flare occurring at 18:30~UT at AR NOAA~11165.
Figures~\ref{fig9}(a--f) show LASCO C2 running difference images of the evolution 
of the CME, which is referred to hereafter as CME1.
After 20:12~UT, another, faster CME (CME2) with a velocity of 700~km~s$^{-1}$ 
appears and follows nearly the same traveling path as that of CME1.
From Figures~\ref{fig9}(d--f), which show CME2, it is seen that it catches up to and 
interacts with CME1 in an act of `cannibalism' \citep{Gopal2001}).
Figure~\ref{fig9}(g) shows a time-distance diagram of both CMEs and 
the EUV coronal loop (Loop2).
CME2 appears to be associated with the flare starting at 19:45~UT on
the same AR.

%%%%%%%%%%%%%%%%%%%%%%%%%%%%%%%%%%%%%%%
%%%%%%%%%%%%%%%%%%%%%%%%%%%%%%%%%%%%%%%

%%%%%%%%%%%%%%%%%%%%%%%%%%%%%%%%%%%%%%%%%%%%%%%%%%%%%%%%%%%%%%%%%%%%%%
\section{Summary and Discussion}

In this study, our goal was to investigate the initial phases of CME-related filament eruptions and
to detect signatures preceding the launch of a CME.
We performed multi-wavelength analyses of the temporal behaviors of the prominence eruption, 
coronal loop expansion, and CME associated with the M4.4 flare that occurred on 
2011 March 8 in AR NOAA 11165.
The 3D velocity field of the erupted prominence was derived using the H$\alpha$ data taken 
by FMT-Peru.
The H$\alpha$ prominence accelerated when high amounts of magnetic energy were released, 
and nonthermal bursts were seen in the microwave (RSTN) and HXR ({\it RHESSI}) from 18:12
to 18:14~UT, after which the prominence rapidly decelerated in the vertical direction and 
changed its Doppler velocity direction.
Finally, the prominence faded out.
This temporal evolution indicates that the cool material seen in the H$\alpha$ line interacted 
with the overlying magnetic environment and failed to eject into space.
%%%
In AIA 304~{\AA} images we can see material falling from there.
Therefore, the disappearance of the H$\alpha$ prominence is caused due to decrease of plasma density.
The change of direction of the erupted prominence was also captured as a deflection of 
bright blobs in {\it STEREO-A}/EUVI 171~{\AA} images, which represents another feature 
of a confined energy release.

Another energy release associated with the nonthermal emissions occurred just 
after the confined event (from 18:19 to 18:20~UT).
This energy release caused an expansion of coronal loops seen in the {\it SDO}/AIA 94~{\AA} 
images (Loop2) and led to a faint CME (CME1).
Hours later, another gradual flare and CME (CME2) occurred.
Using {\it SDO}/AIA images, \citet{Su} investigated in detail the temporal evolution of 
the EUV coronal loops associated with the flare in question (i.e., the flare at around 18:10~UT) 
and the ones associated with another energy release at around 19:45~UT, and interpret 
the two flares as two stages of a single associated event.

By contrast, we found that the first flare comprised further two-stage energy releases and furthermore, 
that the initial energy release represented by the H$\alpha$ prominence eruption appears to be 
confined.
Despite this confinement, the energy release triggered another magnetic reconnection followed by 
the coronal loop expansion and the CME.
As \citet{Kliem2020} and \citet{Gou2019} reported, even a confined flare can cause the eruption of an unstable flux rope.
In our case, we could not confirm the formation of a flux rope as we were observing a limb flare.
Nevertheless, it is possible that dynamic disturbances of the plasmoids and magnetic field triggered 
further energy release during the course of the event, resulting in multi-stage behavior. 
On the other hand, as some authors have reported, erupting filaments are often split into sections 
\citep{Morimoto2003b,Tripathi2006,Guo2010,Cheng2018}.
The blob we observed in H$\alpha$ may be a part of such split filaments and may not be involved 
at all with the flux rope involved in the CME.

%%%%%%%%%%%%%%%%%%%%%%%%%%%%%%%%%%%%%%%

\begin{ack}
The authors acknowledge anonymous referees for their comments and suggestions.
The authors are very grateful with all the staff members of the Kwasan and Hida Observatories,
Kyoto University of Japan, for all the supports and discussions during the FMT-workshops 
and working-group meetings conducted in Japan and in Peru. 
They thank Dr. A. Hillier for his contribution to the FMT-workshops.
They also are grateful to {\it SDO}/AIA and {\it STEREO}/EUVI teams for providing 
high quality data used in this study. 
This work was also supported by the international program ``Climate And Weather of 
the Sun-Earth System - II (CAWSES-II) : Towards Solar Maximum'' sponsored by SCOSTEP. 
This work was also supported by the ``UCHUGAKU'' project of the Unit of Synergetic Studies 
for Space, Kyoto University. 
This work was also supported by JSPS KAKENHI Grant Numbers 25287039, 26400235, 
15K17772, and 16H03955.
A.A. was supported by a Shiseido Female Researcher Science Grant.
The authors would like to thank Enago for the English language review.
\end{ack}

%%%%%%%%%%%%%%%%%%%%%%%%%%%%%%%%%%%%%%%

\end{document}